\documentclass[doublespacing, 12 point]{article}
\usepackage[left=1in, right=1in, top=1in, bottom=1in]{geometry}
\usepackage{amsfonts, amsmath, amssymb}
\usepackage{upgreek}
\usepackage{lineno}
\usepackage{graphicx}
\usepackage[usenames, dvipsnames]{color}
\usepackage[allcolors=blue]{hyperref}
\hypersetup{colorlinks,bookmarksnumbered}

\usepackage[normalem]{ulem}
\usepackage{setspace}
\usepackage{enumerate}
\usepackage{hyphenat}

\usepackage{array}
\usepackage{nicefrac}

\usepackage{multirow}
\usepackage{longtable}
\usepackage{booktabs}
\usepackage{makecell}

\newcolumntype{M}[1]{>{\centering\arraybackslash}m{#1}}

\usepackage{lipsum}

\newcommand{\pr}{$P_{R}$ }
\newcommand{\pfr}{$P_{FR}$ }

\begin{document}
\title{Coherent Control of Optogenetic Switching by Stimulated Depletion Quenching}
\author{Zachary Quine$^1$, Alexei Goun$^1$, Karl Gerhardt, $^2$, Jeffrey Tabor$^2$,  Herschel Rabitz$^1$}

\date{%
    $^1$Department of Chemistry, Princeton University\\%
    $^2$Department of Bioengineering, Rice University\\[2ex]%
    \today
}

%
%

\maketitle

\begin{abstract}
Optogenetics is a revolutionary new field of biotechnology, achieving optical control over biological functions in living cells by genetically inserting light sensitive proteins into cellular signaling pathways.
Applications of optogenetic switches are expanding rapidly, but the technique is hampered by spectral cross-talk: the broad absorption spectra of compatible biochemical chromophores limits the number of switches that can be independently controlled and restricts the dynamic range of each switch.
In the present work we develop and implement a non-linear optical photoswitching capability, Stimulated Depletion Quenching (SDQ), is used to overcome spectral cross-talk by exploiting the molecules' unique dynamic response to ultrashort laser pulses.
SDQ is employed to enhance the control of Cph8, a photo-reversible phytochrome based optogenetic switch designed to control gene expression in E. Coli bacteria.
The Cph8 switch can not be fully converted to it's biologically inactive state ($P_{FR}$) by linear photos-witching, as spectral cross-talk causes a reverse photoswitching reaction to revert to it back to the active state ($P_{R}$). SDQ selectively halts this reverse reaction while allowing the forward reaction to proceed.
The results of this proof of concept experiment lay the foundation for future experiments that will use optimal pulse shaping to further enhance control of Cph8 and enable simultaneous, multiplexed control of multiple optogenetic switches.
\end{abstract}


\section{Introduction and Background}\label{Sec:OG:IntroBG}

\subsection{Introduction}\label{Sec:OG:Introduction}


The variety and complexity of tasks performed by living cells is astonishing: they harvest and refine chemicals from their environment \cite{Albertsch2}, synthesize complex chemicals in response to demand \cite{Gething1992}, respond to changes in their environments in ways that mirror logic and memory \cite{HOPFIELD199453}, self-replicate and multiply under globally limited constraints \cite{AlbertsEssentials,Ardissone2016}, and work together - communicating and combining to form multi-cellular tissues \cite{Albertsch19}.
A cell accomplishes all these tasks by responding to an array of internal and external environmental conditions using various sensor protein complexes and performing a specific biological function, as dictated by a set of instructions in its genetic code. In this way, cells behave as functional machines, performing actions in direct response to stimuli.
Recently, biologists have developed the capability to genetically engineer cells, supplanting the natural sensors with light sensitive proteins. These engineered proteins enable the activation or suppression of specific cellular functions with optical signals, turning a piece of the naturally developed sensing-response system into an externally accessible ``control panel''.
This approach, called \emph{optogenetics} \cite{Deisseroth2011, Pastrana2011}, enables the precise and reversible control of specific biological processes in living cells by external optical stimuli to systematically perturb the system and observe its response\cite{Fenno2011}.

Currently available optogenetic tools can control a range of biological processes including ion channel activation
(enabling high fidelity neuromodulation \cite{Zemelman2002, Boyden2005, XiangLi2005}), organism mobilization \cite{Wen2012, Leifer2011, LeiferPhD2011} and behavior \cite{Montgomery2015, Gunaydin20141535, Husson2013},
protein synthesis in cells \cite{Olson2012,Tabor2011},
%
gene expression or inhibition \cite{Shimizu-Sato2002, Konermann2013, Cao2013, Tyszkiewicz2008},
protein localization at specific cell sites \cite{Levskaya2009, Toettcher20131422, Toettcher2011, Toettcher2011409, Leung02092008},
or sequestration of active proteins away from their site of reaction \cite{Lee2014, Yang01082013}.
Optical techniques can be very appealing for the control and monitoring of biological systems, with many advantages over physical or chemical/pharmacological activation.
Light is a noninvasive tool that interacts strongly with specific biologically relevant molecules and can be applied with high spatial and temporal resolution.
Precise activation signals can be encoded into many controllable parameters of the optical signal, including the light's intensity, frequency spectrum, phase structure, and the duration or temporal pattern of the exposure. Moreover, optical stimuli can be rapidly halted, in contrast to diffusionally controlled chemicals.

In spite of these favorable features, the practical implementation of optogenetics has yet to reaching its fullest potential by spectral cross-talk: the overlapping optical responses of the switches.
Most chromophores, (i.e. the molecules within the optogenetic switch that absorb light) have broad, nearly featureless absorption spectra, which limits the number of switches that can be independently addressed by linear photoexcitation.
For many switches even the absorption spectrum of the activated and de-activated states overlap, preventing the switch from being completely converted from one state to the other.
This inability to distinguish the two states of the switch limits the accessible dynamic range of the system and fidelity of control.

The capability now exists to overcome the spectral cross-talk, as it has been shown that even molecules with nearly identical static absorption spectra can exhibit unique dynamical responses to ultrashort laser pulses \cite{Courvoisier2006}.
These distinguishable dynamics can be accentuated by a set of optimal optical control pulses to discriminate similar molecular species, and this technique is known as Optimal Dynamic Discrimination (ODD) \cite{Li2002, Turinici2004, Li2005,Roth2009, Petersen2010, Rondi2011, Roslund2011, Rondi2012,Goun2016}.
By interacting with the time-dependent non-linear optical response of a molecule, rather than just its static absorption spectrum, we increase the number of control parameters that can be varied to search for orthogonal optical responses that are not restricted by the same spectral cross-talk as in the linear optical regime.
A suitable set of orthogonal optical controls should be capable of separating the optical responses of several similar optogenetic switches that would be extremely difficult to independently address by conventional linear photoexcitation.
The ultimate goal of this research is the simultaneous (multiplexed) control of multiple optogenetic switches over their maximum dynamic range, thereby independently and simultaneously controlling a large number of cellular functions.
In the present work we take the first step toward the goal of multiplexed control of multiple optogenetic switches by achieving enhanced control over a single optogenetic switch, increasing it's dynamic range beyond the photoequilibrium limit imposed by spectral cross-talk.

\begin{figure}
  \centering
  \includegraphics[width=5.9in]{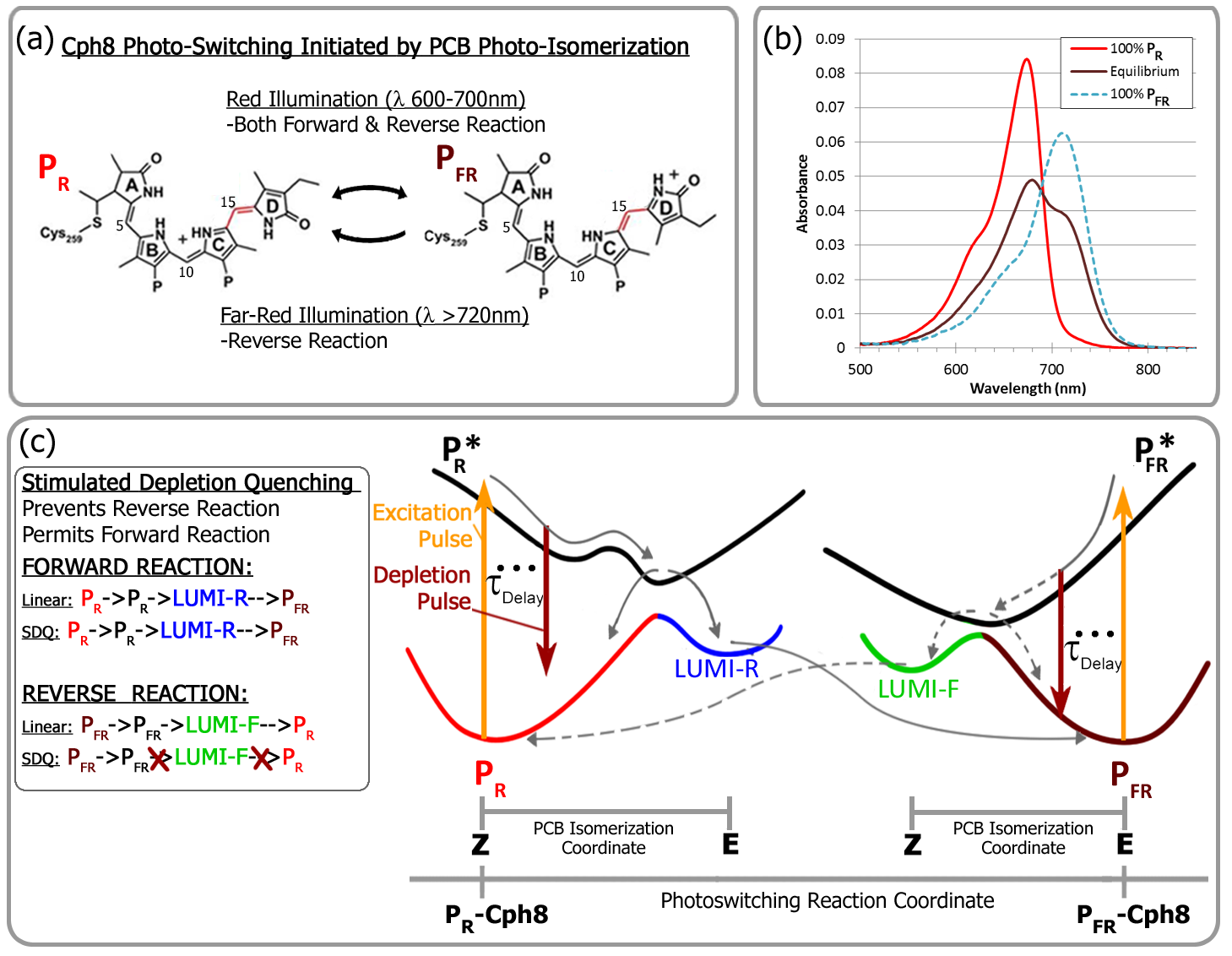}
  \caption[Overview of Experimental Concept]{
  	Overview of the controlled optogenetics experimental concept: (a) Photoswitching of Cph8 is initiated by photoisomerization of the PCB chromophore between an active, red-light absorbing state $(P_{R})$ and an in-active, far-red-light absorbing state $(P_{FR})$. The \pfr state can be completely removed with far-red light, however spectral cross-talk prevents complete removal of the \pr state. 
  	(b) Absorption spectra of a sample of Cph8 in the pure 100\% \pr state, the photoequilibrium mixed state of 65\% $P_{FR}$/35\% $P_{R}$, and the pure 100\% \pfr state that we aim to reach by Stimulated Depletion Quenching (SDQ) enhancement.
  	(c)\emph{The SDQ Concept:} The Cph8 sample in a mix of \pr and \pfr states is exposed to ultrashort excitation and depletion pulses, separated by a delay $\uptau$. Both states are excited and undergo dissimilar coherent ultrafast excited state dynamics along a chromophore isomerization coordinate to either the transient product states (LUMI-R and LUMI-F) or back to their initial ground states (grey arrows). The transient products more slowly transform the remaining protein domains downstream of the chromophore, completing the photoswitching reaction.
  	Precision timing of the depletion pulse permits the selective transfer of a larger portion of the $P_{FR}$* molecules back to the $P_{FR}$ ground state (i.e., the quenched molecules no longer follow grey dashed arrows) compared to the same process for $P_R$*, thereby quenching the reverse reaction more than the forward. This process shifts the equilibrium of the photoswitching reaction toward \pfr, increasing the dynamic range of the Cph8 switch.
}
  \label{OG:FullExpConcept}
\end{figure}

This study utilizes non-linear optical interactions to control the photoswitching reaction of Cph8, a phytochrome based red/far-red sensing optogenetic switch developed for controlling gene expression in E. Coli bacteria \cite{Levskaya2005, Tabor2009,Tabor2011}.
Cph8 is an artificial ``fusion'' protein, a synthesis of the light sensing domains of the Cph1 protein originally found in Cyanobacteria with the Histidine-Kinase Two-Component Switch (TCS) signalling domains of EnvZ, a protein native to E. Coli that is associated with gene transcription \cite{Schmidl2014}.
The chromophore in the Cph8 switch, phycocyanobilin (PCB), transitions between two states by photoisomerization, as shown in Figure~\ref{OG:FullExpConcept}(a).
The gene-expressing ``On'' state absorbs red light and is called \pr, while the biologically inactive ``Off'' state is called \pfr because its absorption spectrum is shifted to longer wavelengths in the far-red. The absorption spectra of these two states is shown in Figure~\ref{OG:FullExpConcept}(b).
The Cph8 switch can be fully converted from the \pfr state to the \pr state by continuous illumination with far-red light ($\lambda \geq$ 720 nm), which is described as the ``reverse'' photoisomerization reaction.
Deactivation of the switch, converting from the \pr state to the \pfr state, is described as the ``forward'' photoisomerization reaction, and due to the overlapping absorption spectra of the two states there is no wavelength of light that will initiate the forward photoisomerization reaction without also initiating the reverse reaction.
The simultaneous photoreaction between the two states under red-illumination comes to photoequilibrium with a maximum \pfr population of $\sim65\%$ (\pr$\sim35\%$). The absorption spectrum of a sample in this equilibrium mixed state is also shown in Figure~\ref{OG:FullExpConcept}(b).
Using ultrashort laser light pulses we will exploit the unique dynamic optical responses of the \pr and \pfr states to selectively ``quench'' the reverse photoswitching reaction, $P_{FR} \rightarrow P_R$, by stimulated depletion, shifting the photoequilibrium to increase the dynamic range of the switch.

A schematic of the proposed \emph{Stimulated Depletion Quenching (SDQ)} mechanism is shown in Figure~\ref{OG:FullExpConcept}(c). The actuation of the Cph8 switch is initiated when the PCB chromophore absorbs a red or far-red photon.

This excitation weakens the bond order of the methine bridge joining the outer C and D rings of PCB, resulting in molecular isomerization by rotation of the D-ring, producing an irreversible cascade of down-stream protein domain reorganizations converting the single electronic excitation and PCB structural change into the actuation of a cellular signaling pathway.
Due to spectral cross-talk of the \pr and \pfr states both are excited by photons of red light, it is not possible to prevent these unwanted electronic excitations of the \pfr state of the PCB chromophore.
However, it \emph{is} possible to prevent the switching of the remainder of the Cph8 protein complex and subsequent activation of the cellular signaling pathways by de-exciting the \pfr state of chromophore before the larger protein ``notices'' that it has been triggered.
After the excitation to the $P_{R}^\ast$ or $P_{FR}^\ast$ state there is a brief interval of time before the chromophore begins isomerization. During this interval the molecules in the $P_{FR}^\ast$ state can be returned to its initial  state $P_{Fr}$ by stimulated depletion, making it appear as though no absorption had occurred and quenching the reverse photoswitching reaction.
Further, to facilitate this capability, there should be sufficient differences in the excited state dynamics and stimulated emission spectra of the two forms of the chromophore to preferentially depopulate the $P_{FR}$* excited state over that of the $P_R$* excited state, allowing the forward reaction to proceed while quenching the reverse reaction.
This selective quenching aims to shift the photoequilibrium of the simultaneous forward and reverse photoswitching reactions, thereby increasing the maximum attainable $P_{FR}$ population.
The total photoswitching reaction can be represented by a simple linear transformation map for an incremental $P_R \rightarrow P_{FR}$ step per laser interaction; upon repeated iteration of the mapping transform (i.e., repeated laser exposure) the full transition of large populations of the Cph8 switch from the \pr to the \pfr state should be possible to accomplish \cite{QuineTheory2018}. The components of the transformation map can be calculated from a small number of measurements of the system, enabling us to rapidly calculate the photoequilibrium population and the enhancement of the dynamic range accessible by the Cph8 switch brought on by the SDQ mechanism. This paper will experimentally determine the map, and then off-line show its iterating capability. It remains for further work to demonstrate the iteration process fully on-line, in the laboratory.



\subsection{Stimulated Depletion Quenching Mechanism}\label{Sec:OG:ExpDesc}

The sample is exposed to two ultrashort laser pulses: an excitation pulse and a depletion pulse.
The spectrum of the excitation pulse overlaps with the absorption spectra of both the $P_{R}$ and $P_{FR}$ states of the PCB chromophore. Exciting the molecule with the ultrashort pulse creates corresponding coherent wave packets from superpositions of vibrational levels on the electronic excited state surfaces of both forms $P_{R}^\ast$ and $P_{Fr}^\ast$ of the chromophore.
These wave packets undergo dissimilar coherent dynamics as they relax from their initial Frank-Condon excitation states toward a coherent vibronic transition through a conical intersection either back to their initial ground state or to a transient intermediate LUMI-(R or F) photoproduct state (see Figure 1).
Before the molecules complete these dynamics, a second pulse arrives after a short, controlled delay ($\uptau_{Delay}$).
This depletion pulse has a central wavelength set to overlap more favorably with the stimulated emission spectrum of the $P_{FR}$* state than the $P_{R}$* state.
The combination of spectral overlap and timing aim to permit the depletion pulse to selectively drive a larger portion of the $P_{FR}$* excited molecules back to the $P_{FR}$ ground state, preferentially slowing the reverse photoisomerization reaction while allowing the forward reaction to continue transferring some portion of the \pr population to the \pfr state.
Those molecules remaining in either of the excited states after the depletion pulse continue their unperturbed dynamics, either isomerizing to the transient photoproduct state or returning to their initial ground state.

 In this work, the excitation and depletion pulse parameters will be varied to determine the wavelengths, powers, pulse durations, and delay timings which selectively deplete the $P_{FR}$* level most effectively, quenching the reverse reaction while minimally hindering the forward reaction.
This exploration of the experimental control parameter space aims to be a feasibility study for the proposed mechanism, identifying a path through the vast, unexplored control landscape toward a region amenable to effective control of the photoswitching reaction, thereby laying the foundation for more sophisticated optimal control experiments on optogenetic switches in the future.

%


\section{Experimental Control of photoswitching by Stimulated Depletion Quenching (SDQ)}\label{Sec:OG:SDQExp}

Here we experimentally utilize the SDQ technique to enhance the control of the Cph8 photoswitching reaction.
These experiments determine the Optical Transformation Matrix (OTM) iterative map associated with a particular laser control parameter set by measuring a group of \emph{characteristic coefficients}, the analog of the same process carried out in simulations \cite{QuineTheory2018}, to calculate the final photoequilibrium populations after many of exposures to the associated laser control pulses.
Further, we characterize the dependence of the SDQ enhancement on the experimentally variable laser control parameters and determining the most effective exposure conditions for maximizing the dynamic range of the switch.
The results of these experiments, presented in Section~\ref{Sec:OG:Results}, are in qualitative agreement with the results of the rate equation model (REM) simulations of the \cite{QuineTheory2018}, and confirm the feasibility of the optimal control of the photoswitching reaction in Cph8 by the SDQ mechanism, laying the ground work for further experiments incorporating pulse shaping and closed-loop optimization to achieve higher fidelity control of optogenetic switches, ideally operating coherently.


\subsection{Experimental Apparatus} \label{Sec:OG:Apparatus}

\begin{figure}
  \centering
  \includegraphics[width=\linewidth]{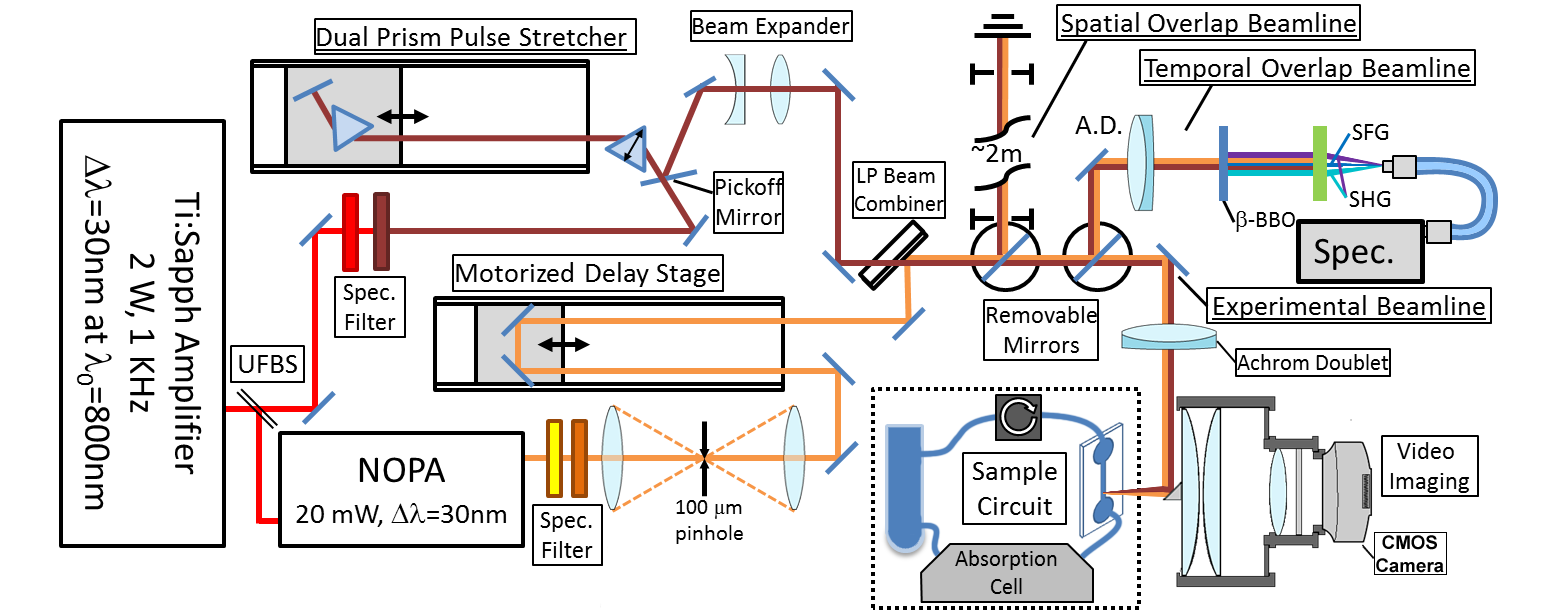}
  \caption[Experimental Schematic]{Schematic of experimental setup including light sources for excitation and depletion laser pulses, beam routing and pulse conditioning optics, sample flow circuit (dotted box, see detailed diagram in Figure~\ref{Fig:OG:ApparatusFluidCircuit}), and systems for aligning, synchronizing, characterizing, and observing beams.}\label{Fig:OG:ApparatusLasers}
\end{figure}
\begin{figure}
  \centering
  \includegraphics[width=.85\linewidth]{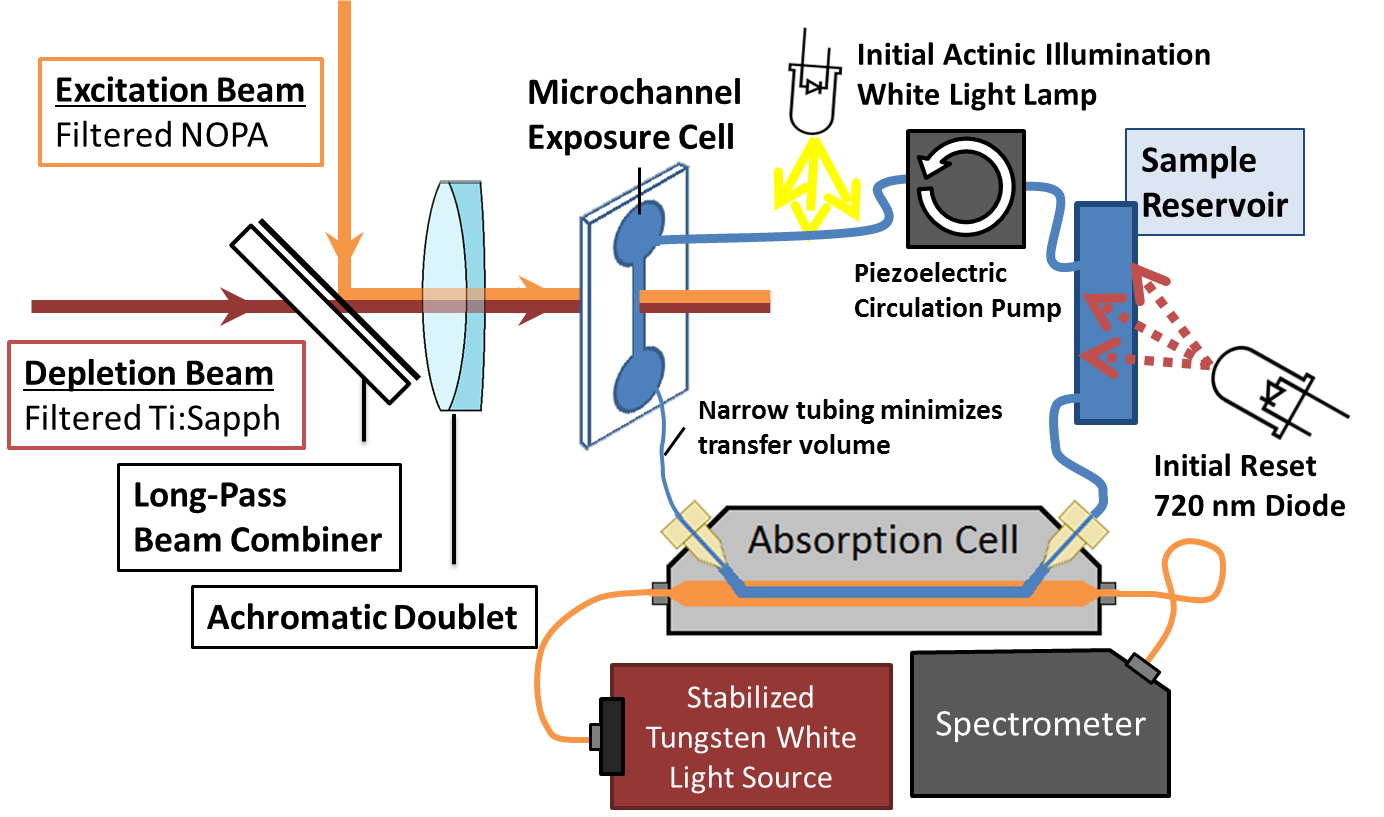}
  \caption[Detailed View of Apparatus Sample Flow Circuit System]{Diagram of SFC which circulates the solution containing the Cph8 protein. Laser exposure initiating the photoisomerization reaction takes place in the microchannel exposure cell, a 100 $\mu$m wide microfluidic channel that flows the sample through the laser focus. Measurement of photoreaction product yield is done in the absorption cell, a fluid filled optical waveguide allowing for long path (5 cm) measurements of low volume (12 $\mu$L) samples. Further description of system provided in the text.} \label{Fig:OG:ApparatusFluidCircuit}
\end{figure}

A diagram of the optical system for the generation, conditioning, and routing of the excitation and depletion pulses to the Sample Flow Circuit (SFC), where the Cph8 sample is stored and circulated for exposure and measurement, are presented in Figure~\ref{Fig:OG:ApparatusLasers}. A more detailed diagram of the SFC is shown later in Figure~\ref{Fig:OG:ApparatusFluidCircuit}. The primary laser source for this experiment is a Ti:Sapphire regenerative amplifier (Coherent Legend) generating high intensity, ultrashort-pulse 800 nm laser pulses of duration $\sim$35 fs (50 nm spectral bandwidth) with average pulse energy 2.1 mJ and 1KHz repetition rate. This source beam is split evenly by a high damage threshold, low GVD 50:50 beam splitter (i.e. UFBS in Figure~\ref{Fig:OG:ApparatusLasers} ) to generate the excitation and depletion beams.

The excitation beam is generated in a Topaz-White Non-collinear Optical Parametric Amplifier (NOPA), which is configured to generate pulses with energy of $\sim$20 $\mu$J/pulse and spectral range from 550-780 nm, peaking at 580 nm. The NOPA beam is filtered with a 610 nm long-pass glass absorption filter (i.e. Spec. Filter in Figure~\ref{Fig:OG:ApparatusLasers} ) and a 650nm short-pass interference filter angle tuned to give a maximum wavelength of 640 nm, producing a nearly constant intensity spectrum from 610 nm to 640 nm. This filtered beam is temporally compressed at the sample exposure site (after passing through all routing optics) by maximizing SHG bandwidth and intensity in a thin 100 $\mu$m thick $\beta$-BBO crystal (not shown in Figure~\ref{Fig:OG:ApparatusLasers} ). The pulse is compressed by adjusting an internal Bi-Prism compressor in the TOPAS–White NOPA.

The depletion beam is generated by spectrally filtering a portion of the Ti:Sapphire amplifier beam using short-pass and long-pass interference filters.
The depletion quenching was analyzed at two wavelengths: 835 nm, corresponding to the peak depletion wavelength predicted by the rate equation model simulations in \cite{QuineTheory2018}, and 775 nm, corresponding to the shortest useable wavelength of the Ti:Sapphire amplifier.
The longer wavelength depletion beam was generated using a 860 nm 40 nm wide band-pass interference filter angle-tuned to pick out the red edge of the Ti:Sapphire amplifier spectrum, producing a pulse with spectral width of 10 nm centered at 835 nm. The shorter wavelength depletion beam was similarly generated with with a 780 nm, 10 nm wide band-pass interference filter angle-tuned to be centered at 775 nm.
The filtered depletion beam was then sent through a dual prism stretcher to adjust the pulse length. The depletion pulse is stretched to enhance the stimulated depletion transition efficiency.

To ensure that the depletion spot size at the sample is much larger than the excitation spot (ensuring that the entire excited sample is uniformly exposed to the depletion pulse) there is an adjustable beam reducing telescope near the beam combiner in the depletion beam line. This telescope can be adjusted while monitoring the overlapped spots at the sample on a camera imaging the exposure spots to confirm proper coverage.

Both beams pass through adjustable delay lines to synchronize the pulses. A computer controlled micrometer driven delay stage is placed in the excitation beam line to introduce a precise delay between the excitation and depletion pulses for time-resolved studies. 

The two beams are independently attenuated with neutral density filters to set the exposure intensity before being spatially combined on an 800 nm long-pass interference filter at 45$\deg$ to transmit the depletion beam and reflect the excitation beam, using close positioned steering mirrors to accurately align the overlap and direction of each beam.
After the beams are combined they can be focused into the microchannel exposure cell. Beams were syncronized and compressed by utilizing SFG and SHG processes with a 100 micron thick BBO crystal. Once synchronized by SFG, the excitation-depletion delay is set by the computer controlled delay stage.

The excitation and quenching beams are focused into the exposure cell by an achromatic doublet with focal length 150 mm. The focusing beams are reflected off a small silver coated turning prism mounted at the center of a 2 inch diameter high numerical aperture (NA) lens before the exposure cell. This lens images the microchannel cell and laser spots onto a video imaging system which monitors the sample and can be used for fine alignment of relative geometry of laser beams with the microfluidic channel. The combined beams are focused onto the exposure cell and the cell is positioned so that the excitation beam diameter is approximately 50\% larger than the microchannel flow cell width and the depletion beam diameter is approximately 50\% larger than the excitation spot. 

A more detailed diagram of the Sample Flow Circuit (SFC) is presented in Figure~\ref{Fig:OG:ApparatusFluidCircuit}. The protein sample circulates in a minimal volume continuously flowing circuit to ensure that each measurement is made on a fresh sample and to mitigate sample photodamage. The SFC is driven by a piezoelectric micro-pump (Takasago fluidics) and consists of a glass reservoir where the bulk of the sample is held, a microchannel exposure cell where the laser activated photoswitching occurs, and a fluid filled optical fiber absorption cell (WPI LWCC-M-50) where the final yield of the photoreaction (degree of Cph8 switching between $P_{R}$ and $P_{FR}$ states) can be measured, all connected by narrow PTFE tubing  to minimize dead volume.

The reservoir is under continuous far-red illumination by a 'reset diode' emitting 730 nm light (M730D2, Thorlabs) which maintains the Cph8 protein in the sample solution in an initial $P_{R}$ state.
Additionally a white light source illuminated a transparent section of tubing connecting the micro-pump to the exposure cell. This light source can be turned on to pre-switch a portion of the molecules before they reach the exposure cell, initializing the sample in a mixture of the $P_{R}$ and $P_{FR}$ states. This initial partition of $P_{FR}$ can be variably set to give the input (\pr, \pfr) population up to $\sim (55,45)\%$.


The exposure cell is a 100 $\mu$m wide 300 $\mu$m deep microchannel flow cell (Translume, Inc.). This narrow channel concentrates the sample into the small area of the laser focal spot, ensuring that the excitation beam completely covers the channel cross-section while still being focused tightly enough to drive the necessary optical transitions.
This exposes the entire flowing volume to a consistent exposure dose set by the laser intensity and fluid flow rate.

The long path absorption cell is a micro-fluidic waveguide capillary cell (LWCC-M-50, WPI inc) that allows highly precise measurements of absorption spectra in low volumes of sample by using the analyzed liquid sample as the internal media in a liquid core optical fiber, keeping the light in contact with the sample by total internal reflection at the liquid-core interface and enabling long optical path lengths. The total internal volume of the absorption cell is 12 $\mu$L.
The sample is probed with a stabilized Tungsten white light lamp (Thorlabs SLS202) that provides a smooth, continuous spectrum well-suited to absorption measurements. The light source is stabilized by an electronic feedback circuit to provide $<$0.1\% optical power drift per hour for long term stability. The light source is coupled to the LWCC by a fiber optic cable, and the transmitted probe is coupled to a compact spectrometer (Ocean Optics MAYA4000) by another fiber optic cable. 


\subsection{Data Collection and Analysis}\label{Sec:OG:DataProc}


Using the apparatus described in the last section, the sample is (i) circulated as absorption spectra are continuously measured, (ii) the absolute concentration of the two conformational states of Cph8 are extracted, (iii) saved, and (iv) indexed by their time of acquisition. As the laser exposure conditions are varied, the collected absorption spectra and relative photoswitching reaction product populations change in response, as shown later in Figure~\ref{R1}. By varying the experimental laser control parameters and measuring the final photoswitching reaction product states under a series of six \emph{Exposure Conditions} (described in Table~\ref{Tab:OG:ExpCond}) from the final product state populations it is possible to calculate a set of six \emph{characteristic coefficients} (described in Table~\ref{Tab:CharCoeff}). These \emph{characteristic coefficients} are a series of reaction coefficients that quantify the effects of the excitation and depletion pulses in ways that facilitate analysis and allow us to build the OTM map to calculate the enhanced photoequilibrium populations.

\begin{figure}
  \centering
  \includegraphics[width=.9\linewidth]{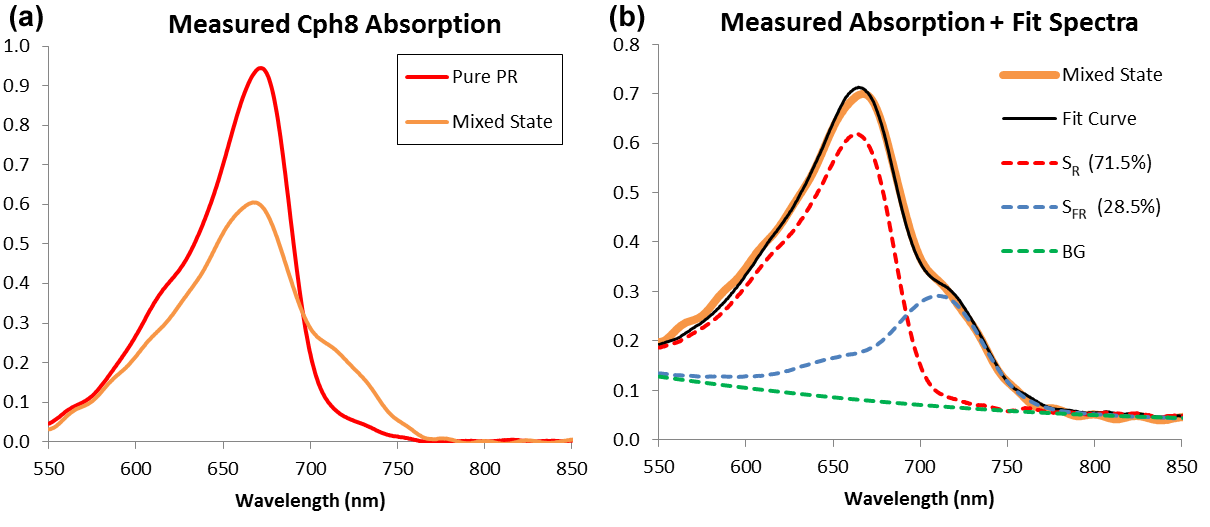}
  \caption[Representative Absorption Spectra from Cph8]{(a) Representative absorption spectra of Cph8 taken in the absorption cell under typical experimental concentrations and conditions. This graph contains spectra from Cph8 in a `pure' \pr state and a mixed state of 71.5\% $P_{R}$ and 28.5\% $P_{FR}$. (b) The mixed state spectrum is fit to a linear combination of the reference spectra and a nonlinear background (BG). Dashed traces show the relative contribution of the \pr and \pfr states to the total absorption fit curve (black trace). }\label{Fig:OG:AbsSpec}
\end{figure}

Representative absorption measurements from two samples are shown in Figure~\ref{Fig:OG:AbsSpec}(a), while Figure~\ref{Fig:OG:AbsSpec}(b) shows the fitting of one of these absorption measurements to known reference spectra to extract the concentrations of the component states present in the sample.
The two samples shown in Figure~\ref{Fig:OG:AbsSpec}(a) include a sample in the pure \pr state (red trace) under continuous illumination at 730 nm as well as a sample with a mixed population containing Cph8 in both \pr and \pfr states (orange trace) produced by white light illumination.
Reset samples in the `pure' \pr state typically retain a small residual population of $1-3\%$ \pfr.
Backgrounds have been subtracted to place the two spectra on the same baseline.

The measured absorption spectra are fit to a linear combination of absorption peaks and a third order polynomial background:
\begin{equation}\label{Eq:OG:FitFun}
  S(\lambda, c_i)=c_{R}S_{R}(\lambda)+c_{FR}S_{FR}(\lambda)+c_{R'}S_{R'}(\lambda)+c_{W}S_{W}(\lambda)+\sum_{n=0}^{3}c_n\cdot(\lambda-\lambda_0)^n
\end{equation}
\noindent The four absorption peaks are the known reference spectra of Cph8 in the \pr and \pfr states ($S_{R}(\lambda)$ and $S_{FR}(\lambda)$) (measured independently and verified by comparison to ref~\cite{Lamparter2001}), a small water overtone peak at 970 nm ($S_{W}(\lambda)$), and a  blue-shifted sub-population of $P_{R}$-Cph8 with a spectrum peaking at 655 nm that develops over the course of experiments ($S_{R'}(\lambda)$). This sub-population is discussed further below.
The analytical function is fit to the data by an unconstrained least squared residual fitting algorithm.
The reference spectra in Equation~\ref{Eq:OG:FitFun} are scaled to the known extinction coefficients found in the literature \cite{Lamparter2001} and multiplied by the optical path length of the absorption cell (5 cm), so the fit coefficients $c_R$, $c_{R'}$, and $c_{FR}$ correspond to absolute concentrations of the component states of Cph8 in Molar units.
The set of experimentally controlled parameters investigated include: the average power of the excitation and depletion beams ($\langle P_{Ex}\rangle$ and $\langle P_{Dp}\rangle$), the central wavelength and bandwidth of the excitation pulse spectrum (($\lambda_0^{Ex}$,$\Delta\lambda_{Ex}$) and depletion pulse spectrum ($\lambda_0^{Dp},\Delta\lambda_{Dp}$)), the laser pulses' temporal widths ($\Delta t_{Ex}$ and $\Delta t_{Dp}$),
the time delay between the two pulses ($\uptau_{(Ex,Dp)}$), and the sample flow rate (q), which (in conjunction with the laser repetition rate) defines the number of laser pulses the sample is exposed to as it passes through the laser focal spot in the microchannel exposure cell.
This set of experimental control parameters,
\begin{align}\label{exExpContParam}
    \left[\langle P_{Ex}\rangle, \lambda_0^{Ex}, \Delta\lambda_{Ex}, \Delta t_{Ex}, \langle P_{Dp}\rangle, \lambda_0^{Dp},\Delta\lambda_{Dp}, \Delta t_{Dp}, \uptau_{(Ex,Dp)}, q\right],
\end{align}
\noindent defines a single point in the high-dimensional parameter space to manipulate the molecular switch.
The molecular response is related to all of these parameters collectively through coupled, complex relationships.

In \cite{QuineTheory2018}, we define the final product populations of the photoswitching reaction as a linear transformation from the initial state population distribution using a set of \emph{characteristic coefficients}:
\begin{subequations}\label{eqSMFCC}
  \begin{align}
    P_{R,f}={}&\left(1-D_R-(1-Q_R ) Y_{R,FR} \right)P_{R,i} + \left((1-Q_{FR} ) Y_{FR,R} \right)P_{FR,i} \\
    P_{FR,f}={}&\left((1-Q_R ) Y_{R,FR} \right)P_{R,i} + \left(1-D_F-(1-Q_{FR} ) Y_{FR,R} \right)P_{FR,i}
\end{align}
\end{subequations}
The per exposure yield in the forward direction, ${Y_{R,FR}}$, is the fraction of molecules transferred from \pr to \pfr by the excitation pulse alone with no depletion and the forward quenching coefficient, ${Q_R}$, is the relative change in forward yield brought on by the dual pulsed excitation-depletion sequence.
The corresponding coefficients ${Y_{FR,R}}$ and ${Q_{FR}}$ are defined for the reverse reaction in the same way.
${D_{R}}$ and ${D_{FR}}$, associated with the loss of protein within the sample due to photodamage by the high intensity laser pulses. Care was taken to minimize photodamage by limiting peak laser intensities, but it is a loss channel that must be accounted for in the experimental measurements.

To calculate the six unknown coefficients we measure the final product populations under series of six \emph{Exposure Conditions} (described and collected in Table~\ref{Tab:OG:ExpCond}).
The sample can be left to circulate with no laser exposure, measuring the initial population distribution in either the pure-\pr (EC1, $\binom{R_0}{F_0}$) or mixed (EC4, $\binom{R_{M0}}{F_{M0}}$) initial conditions.
From either of these initial conditions the sample can be exposed to the excitation laser alone (EC2, $\binom{R_{Ex}}{F_{Ex}}$;EC5, $\binom{R_{MEx}}{F_{MEx}}$), providing the linear photoswitching reaction products, or the excitation-depletion pair (EC3, $\binom{R_{ED}}{F_{ED}}$;EC6,  $\binom{R_{MED}}{F_{MED}}$), producing the non-linear SDQ-enhanced photoswitching reaction products.
These six \emph{Exposure Conditions} make up a ``\emph{Full Data Set}'', thereby producing sufficient measurements to calculate the full set of \emph{characteristic coefficients} for a specified laser control parameter set.
From the photoswitching reaction products starting in the pure \pr condition we calculate the yield, quenching, and damage coefficients of the forward reaction.
Using the final product states collected starting in the mixed condition and the calculated forward characteristic coefficients, we can calculate the reverse yield, quenching, and damage coefficients.
The set of six \emph{characteristic coefficients} is collected in Table~\ref{Tab:CharCoeff} along with their relation to the measured final product populations under the six \emph{Exposure Conditions}.

\begin{table}
\centering
\footnotesize
	\begin{tabular}{>{\centering\arraybackslash}p{.30in}| @{}>{\centering\arraybackslash}m{.85in}@{}|@{} >{\centering\arraybackslash}m{.85in} @{}| m{3.6in} }
\toprule
\toprule

\textbf{EC\#} &
\makecell{\textbf{Initial} \\ \textbf{State} $(\mu M)$}&
\makecell{\textbf{Final} \\ \textbf{State} $(\mu M)$}&
\textbf{Description of Exposure Condition} \\

\midrule

1& &$\begin{bmatrix} R_{\text{\tiny 0}} \\ F_{\text{\tiny 0}} \\ \end{bmatrix}$&
\uline{Pure $P_R$ state:} Reservoir under continuous 730 nm diode illumination to hold switch in $P_R$ state. (typically $1-3\%$ \pfr remains).\\
2&\makecell{$\begin{bmatrix} R_{0} \\ F_{0} \\ \end{bmatrix}$ } &$\begin{bmatrix} R_{\text{\tiny Ex}} \\ F_{\text{\tiny Ex}} \\ \end{bmatrix}$&
\rule{0pt}{3ex} \uline{Linear Excitation:} Sample exposed to excitation laser alone. Change in $P_{FR}$ gives forward yield ($Y_{R,FR}$), while difference in $P_{FR}$ produced and $P_R$ lost gives forward damage coefficient ($D_R$).\\
3& &$\begin{bmatrix} R_{\text{\tiny ED}} \\ F_{\text{\tiny ED}} \\ \end{bmatrix}$&
\rule{0pt}{3ex} \uline{Excitation+Depletion:} Sample exposed to Excitation and Depletion pulse pair to excite and then quench forward reaction ($Q_R$). Power, spectrum, and timing of pulses are varied to map dependence of $Q_R$.\\
\midrule
4& &$\begin{bmatrix} R_{\text{\tiny M0}} \\ F_{\text{\tiny M0}} \\ \end{bmatrix}$&
\uline{Mixed Initial State:} Dual pre-illumination (730 nm diode+white lamp) prepares sample in a mixed initial state. $[P_{FR}]_{init}$ is dependent on lamp intensity and sample flow rate.\\

5&$\begin{bmatrix} R_{\text{\tiny M0}} \\ F_{\text{\tiny M0}} \\ \end{bmatrix}$&$\begin{bmatrix} R_{\text{\tiny MEx}} \\ F_{\text{\tiny MEx}} \\ \end{bmatrix}$&
\rule{0pt}{3ex} \uline{Mixed Linear Excitation:} Sample with initial $P_{FR}$ population undergoes forward and reverse reaction. Using the calculated $Y_{R,FR}$ the reverse yield ($Y_{FR,R}$) can be extracted.\\

6& &$\begin{bmatrix} R_{\text{\tiny MED}} \\ F_{\text{\tiny MED}} \\ \end{bmatrix}$&
\rule{0pt}{3ex} \uline{Mixed Excitation+Depletion:} Sample undergoing forward and reverse reaction is exposed to the depletion pulse. Using the calculated $Y_{R,FR}$, $Y_{FR,R}$, \& $Q_R$, the Reverse Quenching Coeff. ($Q_{FR}$) can be extracted.\\	

\bottomrule
	\end{tabular}
	\caption[Exposure Conditions]{Exposure conditions to calculate the six Characteristic Coefficients from measured final product states. The six exposure conditions defines a \emph{Full Data Series} for a single control parameter set. The Full Data Series is required to calculate the full set of \emph{Characteristic Coefficients} for that set of control parameters; though, useful information about individual characteristic coefficients can be extracted from a partial data series.}
	\label{Tab:OG:ExpCond}
\end{table}

\begin{table}
\centering
\renewcommand{\arraystretch}{1.5}
\footnotesize
\begin{tabular}{>{\centering\arraybackslash}m{1.2in} >{\centering\arraybackslash}m{0.7in} >{\centering\arraybackslash}m{3in} }
\toprule
\toprule

\textbf{Characteristic\break Coefficients}& \textbf{Symbol} & \nohyphens{\textbf{Calculation from Measured\break Final Product Concentrations}}\\
\midrule
Forward\break Yield&$Y_{R,FR}$&${(F_{Ex}-F_0)}/{R_0}$\\
    \hline
Forward\break Damage&$D_R$&$1-Y_{R,FR}-R_{Ex}/R_0$\\
    \hline
Forward\break Quenching&$Q_R$&$1-(\nicefrac{(F_{ED}-F_0)}{R_0} )/Y_{R,FR}$\\
    \hline
Reverse\break Yield&$Y_{FR,R}$&$\dfrac{R_{MEx}-R_{M0} (1-Y_{R,FR}-D_R )}{F_{M0}}$\\
    \hline
Reverse \break Damage&$D_{FR}$&$1-\dfrac{F_{MEx}+R_{MEx}-(1-D_R)R_{M0}}{F_{M0}}$\\
    \hline
Reverse\break Quenching&$Q_{FR}$&$1- \dfrac{R_{MED}-R_{M0}(1-(1-Q_R )Y_{R,FR}-D_R )}{Y_{FR,R}\cdot F_{M0}}$\\
\bottomrule
		\end{tabular}
	\caption[Characteristic Coefficients]{The Characteristic Coefficients relate the measured
  final photoswitching reaction products to the ultrafast interactions being used to control the process.
  The first and second columns give the name and symbols for each coefficient, while the last column describes the method of calculating the coefficient from the final product state populations measured under the exposure conditions in Table~\ref{Tab:OG:ExpCond}.}
  \label{Tab:CharCoeff}
\end{table}

Once these \emph{characteristic coefficients} are calculated, we can form an Optical Transformation Matrix (OTM), just as we did in Equation 3.

\begin{equation}\label{eq:OG:expOTMccs}
 \begin{bmatrix}
          P_{R,f} \\
          P_{FR,f} \\
        \end{bmatrix} =\begin{bmatrix}
                        (1-(1-Q_{R})Y_{R,FR}) & (1-Q_{FR})Y_{FR,R} \\
                        (1-Q_{R})Y_{R,FR} & (1-(1-Q_{FR})Y_{FR,R})
                      \end{bmatrix} \begin{bmatrix}
          P_{R,i} \\
          P_{FR,i} \\
        \end{bmatrix}
\end{equation}


\noindent The damage coefficients are omitted from the OTM, as the inclusion of the loss term prevents calculation of the steady-state equilibrium solution from the eigensolutions (as no steady-state exists and the sample continuously loses population). Rather, accounting for the damage coefficients in the calculation of the other \emph{characteristic coefficients} ensures their accuracy.
With the experimentally measured OTM we can calculate the eigensolutions of the matrix. The projected photoequilibrium can be obtained from the eigenvector with eigenvalue $\lambda_{eq}=1$.
A \emph{Full Data Set} is defined by measuring the final product concentrations under all six of the exposure conditions in Table~\ref{Tab:OG:ExpCond} for a given control parameter set (see Figure~\ref{R2}).
Each time one of the control parameters in the set is varied a Full Data Set is measured to extract the \emph{characteristic coefficients} at that point in the parameter space landscape.
The \emph{characteristic coefficients} are interdependent, thus making it critical that experimental conditions remain consistent throughout the measurement of each full data series. 

 In Figure~\ref{R1} we present data collected from a portion of a typical measurement series, taken over the course of one hour.
The concentration of the states of Cph8 in the sample are continuously calculated from the measured absorption spectra as the sample circulates under varied laser illumination conditions, designated in the figure by the \emph{Exposure Condition} number corresponding to Table~\ref{Tab:OG:ExpCond}.
The sample is initially in a pure $P_{R}$ configuration, with total Cph8 concentration of 1.62 $\mu$M.
Between 28-43 minutes in the figure the sample is in a mixed initial state $\binom{R_{M0}}{F_{M0}}$ with a non-zero $P_{FR}$ concentration.
Over the course of the measurement series the sample is exposed to the laser sources to initiate the photoswitching reaction by either a linear, single-pulse excitation or a non-linear, dual-pulse excitation-depletion pair.
The periods under excitation-only exposure are highlighted in yellow, while the periods exposed to the excitation-depletion pair are highlighted blue; white regions have no laser exposure. 

In the figure the concentrations of two spectrally distinct sub-populations of the \pr state is denoted as, $R_{670}$ and $R_{655}$, named by the peak of their absorption spectra.
In all measurements the Cph8 sample is observed to be initially free of the $R_{655}$ sup-population, which gradually develops after extended periods circulating in the experimental apparatus.
The conversion to $R_{655}$ of the sample is observed under complete dark circulation with only intermittent observation by the probe to minimize exposure to light, indicating that the effect is not due to photodamage of the protein.
A similar blue-shifted ground state sub-population is observed by Larsen et.al. \cite{Kim20140417}.
The $R_{655}$ sub-population participates in the photoswitching with the same yield as the initial $R_{670}$ population. Typically, we complete the measurement series before significant populations of $R_{655}$ develop.

\begin{figure}
  \centering
  \includegraphics[width=5.5in]{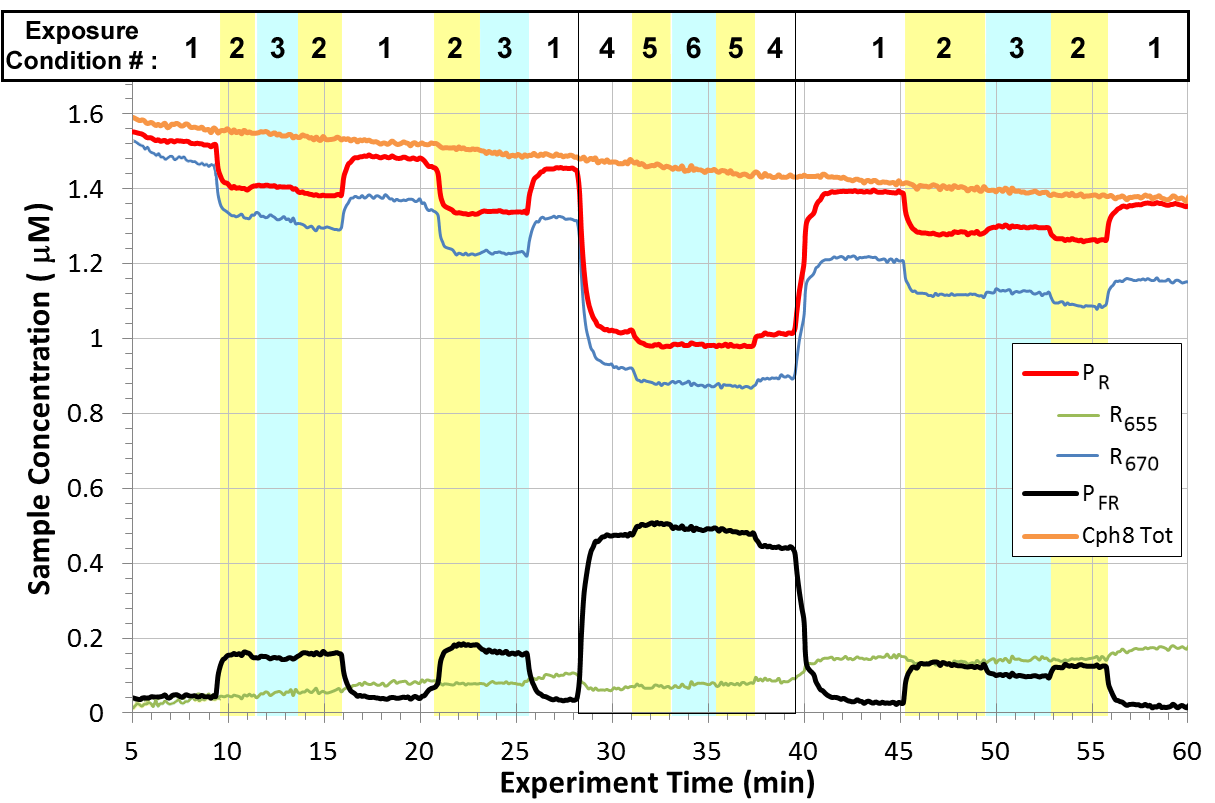}
  \caption[Data from representative measurement series]{Charting of the measured concentration of the $P_{R}$ and $P_{FR}$ states of Cph8 protein over the course of several measurement series under varied exposure conditions, labeled by number corresponding to Table~\ref{Tab:OG:ExpCond}. Regions highlighted yellow are exposed to the excitation pulse alone, regions highlighted blue are exposed to the excitation-depletion pulse combination, and non-highlighted regions have no laser exposure.Two spectrally distinct sub-populations of \pr are charted in the figure, $R_{655}$ and $R_{670}$, discussed in the text.}\label{R1}
\end{figure}

The first feature of note in the figure is the steady decrease of the total concentration of Cph8 protein in the sample (orange trace in Figure~\ref{R1}). Over the course of the measurement series the total protein concentration falls 14\%.
This decline in total protein concentration is seen under all conditions of circulation, including complete darkness, and is unrelated to the laser exposure.
The source and nature of the sample degradation was not investigated in this work, though there are two suspected causes.
The first is aggregation of the protein complexes, which then fall out of solution and are trapped in the sample reservoir, reducing the overall concentration.
The second is denaturation of the protein due to age, handling, or environmental conditions.
The development of the $R_{655}$ sub-species does not appear to be associated with the protein loss, as the two effects develop with a different time dependence and are believed to be unrelated processes.
Regardless of the source of the non-optical protein loss, the trend is linear and consistent across the series of measurements, and can be fit and removed from the measurements without adversely affecting the data.
The removal of this gradual non-optical loss trend is the first step in the processing of a measurement series.

Next, the populations of the $P_{R}$ and $P_{FR}$ states are averaged over a fixed time period under each exposure condition after the measured concentration levels have stabilized. Once a \emph{Full Data Set} has been measured under a given set of experimental control parameters, the \emph{characteristic coefficients} of that parameter set can be calculated and the experimental control parameters can be varied to the next point in the survey.
The collection of a \emph{Full Data Set} for a given laser parameter setting need not be measured consecutively or in any particular order, so long as all six exposure conditions are measured under similar experimental conditions to reliably calculate all six characteristic coefficients and fully describe the photoswitching reaction at that laser parameter setting and predict the final equilibrium population. The order of data collection was alternated to suppress systematic errors.

\begin{figure}
  \centering
  \includegraphics[width=5.5in]{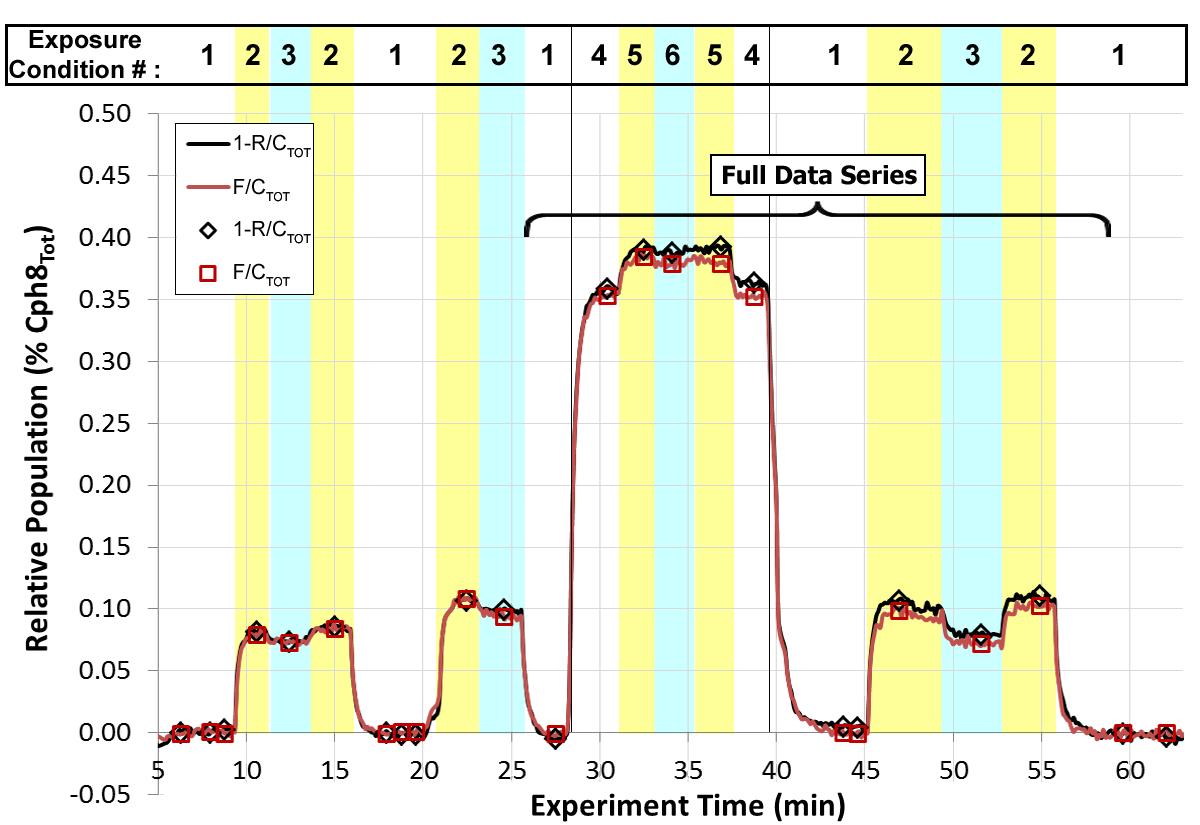}
  \caption[Processed Data from representative measurement series]{Processed measurement series: The relative decrease of \pr, $(1-R/C_{Tot})$, overlaps with the relative increase of \pfr, ($F/C_{Tot}$), where $C_{Tot}$ is the total Cph8 concentration, demonstrating `clean' switching between the two states with little loss to photodamage. As in Figure~\ref{R1}, \emph{Exposure Conditions} are labeled corresponding to Table~\ref{Tab:OG:ExpCond}: yellow regions are exposed to the excitation pulse, blue regions are exposed to the excitation-depletion pulse combination, and non-highlighted regions have no laser exposure. The bracketed region from 27-60 minutes defines one \emph{Full Data Series} collected at $\langle P_{Ex}\rangle\!\!=\!\!600\mu W$, $\langle P_{Dep}\rangle\!\!=\!\!4.5mW$, and $\uptau_{Ex,Dp}\!\!=\!\!0.5ps$ (the full control parameters are given in the text).}\label{R2}
\end{figure}

Processed data from the same measurement series shown in Figure~\ref{R1} are presented in Figure~\ref{R2}, where the non-optical loss trend has been fit and removed and the $P_{R}$ and $P_{FR}$ state populations plotted as ratios of the total Cph8 concentration.
The markers indicate the average state population at each specific exposure condition indexed by the experiment time at which the exposed sample traveled from the exposure cell to the absorption cell and the measured state concentration had stabilized.
The standard deviation of the measured concentration over the averaged interval in each exposure condition is measured and used to calculate the error of each \emph{characteristic coefficient} by propagation of uncertainty. 
In the processed measurement series in Figure 6, the overlap of the \pfr population increase with the \pr population decrease clearly illustrates the low damage coefficients.
The larger net production of $P_{FR}$ per laser exposure at the same power when starting from zero initial concentration (Exposure Condition 2, EC2), compared to when the initial concentration of $P_{FR}$ is higher (EC5) illustrates that in EC5 the forward and reverse photoreactions are taking place simultaneously and the sample is near photoequilibrium.
The regions in Figure~\ref{R2} when the sample was exposed to the excitation-depletion pulse combination are highlighted in blue (EC3 \& EC6).
Because the number of laser pulse iterations during the exposure is kept low to more rapidly scan a larger domain of laser control parameters, the excitation plus Depletion pulse combination causes the \pfr population to \emph{decrease}, rather than increase.
This result is expected of the SDQ mechanism when the number of laser pulse iterations per exposure is less than the number required to surpass the linear excitation.
This occurs because the SDQ mechanism functions by slowing down both the forward and reverse reactions, but slowing the reverse reaction more, enabling a greater final photoequilibrium while requiring more laser pulse iterations to reach it.
Because of the information we extract from the iterative mapping using the OTM we are able to calculate the photoequilibrium of the laser exposure without waiting for this extended period.


The bracketed region between 27-60 minutes encloses a single \emph{Full Data Series }of measurements taken at the following experimental control parameter settings: [$\langle P_{Ex} \rangle, \langle P_{Dep} \rangle,(\lambda_0^{Ex},\Delta\lambda_{Ex} ), (\lambda_0^Dp,\Delta\lambda_{Dp} ), \uptau_{Ex,Dp}$] = [600 $\mu$W, 4.5 mW, (625 nm, 40 nm), (835 nm, 10 nm), 0.50 ps].
The two partial data series at earlier times are taken at different excitation and depletion powers.
The first partial data series is taken at $\langle P_{Ex}\rangle=400\mu W$, $\langle P_{Dep}\rangle=1.5mW$, and $\uptau_{Ex,Dp}=0.33 ps$, and the second partial data series at $\langle P_{Ex}\rangle=600\mu W$, $\langle P_{Dep}\rangle=1.5mW$, and $\uptau_{Ex,Dp}=0.5 ps$.
Though these measurements can not be used to calculate the full set of \emph{characteristic coefficients}, they are useful for showing trends in the power and delay dependence of the forward Yield and forward Quenching coefficients.
From the \emph{Full Data Series} we can calculate the full set of \emph{characteristic coefficients}:
[$Y_{R,FR}, Q_R, D_R$]=[$0.1160\pm0.0035, 0.29\pm0.02, 0.0013\pm0.0006$], and
[$Y_{FR,R}, Q_F, D_F$]=[$0.1010\pm0.003, 0.493\pm0.055, 0.0009\pm0.0009$].
The errors in the calculated \emph{characteristic coefficients} derive from propagation of the standard deviation of the measured sample concentrations through the equations for each coefficient (see Table~\ref{Tab:CharCoeff}).

We see that the forward and reverse reaction Yield coefficients, $Y_{R,FR}$ and $Y_{FR,R}$, have similar magnitude with an advantage to the forward reaction, as expected at this excitation wavelength.
The damage coefficients are very low, which was common throughout the experiments.
As stated earlier, efforts were taken to limit photodamage, and significant levels of damage (i.e., defined as a damage coefficient higher than $\sim 1.5\%$) were only seen in samples compromised by age or environment (stored for several weeks at 4$^\circ$C or overnight at room temperature).
In an extended measurement series lasting $>$10 hours the photodamage parameter would increase over the course of the experiment, but typically remained under a few percent.
The Quenching coefficient of the reverse reaction, $Q_{FR}$, is 1.7 times greater than the forward Quenching coefficient, $Q_R$, a significant enhancement factor.
Using the calculated characteristic coefficients, we construct the OTM and calculate the equilibrium product population from its eigensolutions.
Analytically calculating the errors of the eigensolutions of a matrix from the errors of its components is prohibitively complicated, so the eigensolution errors were numerically computed by statistical sampling. We calculated the eigensolutions of 1000 matrixes with components defined by the measured characteristic coefficients with Gaussian variations of magnitude given by the associated coefficient errors. The standard deviation of these eigensolutions provides the estimated error in the photoequilibrium population.
The projected photoequilibrium threshold of the linear excitation is $[P_{FR}]_{EQ}^{lin} = 0.54\pm0.09$. Based on the Quenching coefficients calculated from the measurements of the full data series in Figure~\ref{R2} the projected SDQ-enhanced equilibrium threshold is $[P_{FR}]_{EQ}^{SDQ} = 0.62\pm 0.10$. This is an increase of the maximum \pfr population by 13\% or a decrease of the minimum  residual \pr population by 21\%, which is a sizeable enhancement.

In the following section we present the calculated \emph{characteristic coefficients} measured over a range of laser parameter settings and discuss the functional dependence of the photoswitching reaction on the experimental control parameters.

 which is

\subsection{Results}\label{Sec:OG:Results}

\begin{figure}
  \centering
  \includegraphics[width=6in]{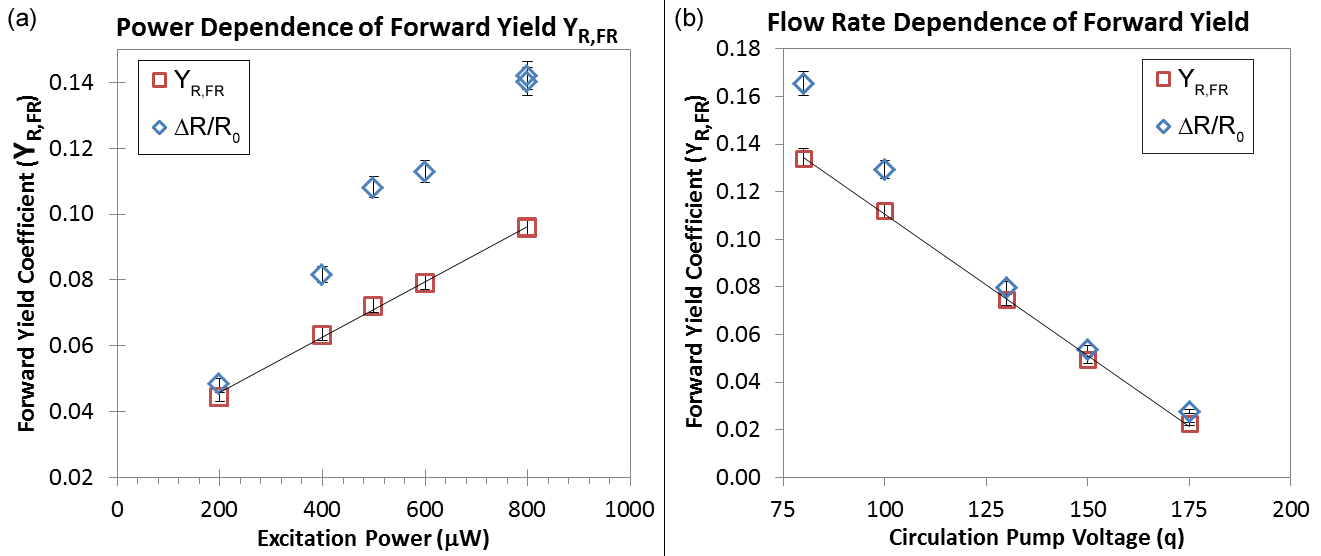}
  \caption[Power and Flow Rate Dependence of Forward Yield]{Power and flow rate dependence of the per-exposure yield of the forward ($P_{R}\rightarrow P_{FR}$) photoswitching reaction: (a) $Y_{R,FR}$ measured as a function of Excitation power. Measurements were taken at high flow rate (q=150)
  to ensure consistent sample characteristics and prevent photodamage at high powers.
  (b) $Y_{R,FR}$ measured at 200 $\mu$W Excitation power while varying the piezoelectric pump driving voltage (directly related to flow rate).
  Difference between $\nicefrac{\Delta R}{R_0}$ and $Y_{R,FR}$ gives the damage coefficient $D_R$, which increases as laser power increases or flow rate decreases and sample is subjected to more laser intensity per exposure. Error bars ($\sim$size of markers) are calculated by propagation of the standard deviation of measured concentrations through the calculation of the Yield coefficients.
  }\label{ExciteDep}
\end{figure}

The forward Yield ($Y_{R,FR}$) is the easiest of the \emph{characteristic coefficients} to calculate from the experimental measurements. It can be computed after measuring just two exposure conditions, requiring only the change in $P_{FR}$ concentration after exposure of the initially pure \pr state to the excitation pulse.
Measurements show that the forward Yield is linearly related to the average power of the excitation beam as well as the sample flow rate over the range of values investigated in these experiments.
Figure~\ref{ExciteDep}(a) shows the linear power dependence of the forward Yield taken at a constant flow rate, while increasing the average power of the excitation beam with a variable neutral density filter.
These measurements were taken at high flow rate (flow rate setting q=150 V) to limit photodamage at higher excitation powers and to collect the data quickly, ensuring consistent experimental conditions over the series.
Figure~\ref{ExciteDep}(b) shows the decrease in the forward Yield as the number of laser pulses per exposure iteration is decreased: the driving voltage of the piezoelectric circulation pump sets the flow rate through the circuit, as the sample flows through the exposure cell more quickly it is exposed to fewer laser pulses.
The flow rate is observed to be directly related to the driving voltage of the piezoelectric pump over this range of pump voltages.
Figure~\ref{ExciteDep} also shows the relative decrease of the \pr state $(\nicefrac{-\Delta R}{R_0})$. The difference between this measure and the forward Yield gives the damage coefficient, $D_R$. At higher power and slower flow rates the increased laser intensity and number of pulses per exposure causes $D_R$ to rise.
Most measurements were taken with lower excitation power, ranging from 100 - 250 $\mu$W, and the flow rate is kept as slow as possible without leading to significant photodamage or impacting the ability to complete a measurement series under consistent sample composition.

\begin{figure}
  \centering
  \includegraphics[width=4in]{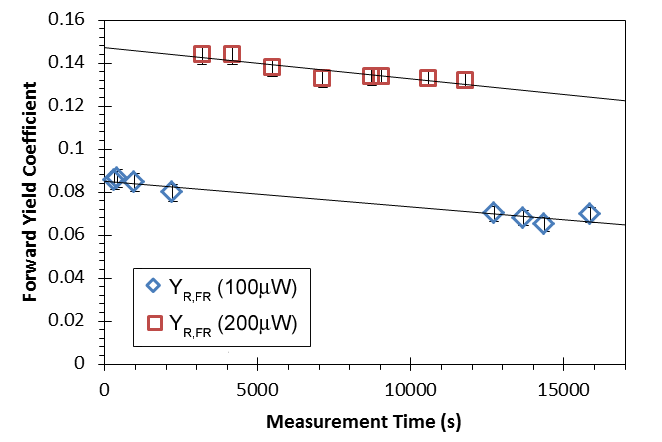}
  \caption[Forward Yield Coefficient]{Gradual decline of forward Yield coefficient over a long measurement series ($\sim$4 hours). Laser power, spectrum, and alignment are constant over the course of measurements; therefore, the decrease in yield is attributed to a change in sample makeup over the course of the measurements.}\label{FYield}
\end{figure}

In Figure~\ref{FYield} we plot the calculated forward Yield from a long series of measurements lasting over four hours and at two excitation powers.
The figure shows the degree of stability of the experimental conditions over the duration of this extended measurement series.
The Yield coefficients in this series of measurements are higher than those shown in Figure~\ref{ExciteDep}(a) because the flow rate is lower (q=100).
There is a general systematic decrease in the $Y_{R,FR}$ value, but the trend is gradual and consistent across the measurement series. 
The power and beam overlap were checked repeatedly during this time to confirm that they remained fixed, therefore the drop in yield must be attributed to a variation in sample composition or flow rate.
The trend is constant in time and was compensated for in the analysis of other \emph{characteristic coefficients} calculated from this measurement series by fitting the measured $Y_{R,FR}$ values to a linear trend and using the estimated forward yield at any measurement time to calculate the other \emph{characteristic coefficients}.

\begin{figure}
  \centering
  \includegraphics[width=4in]{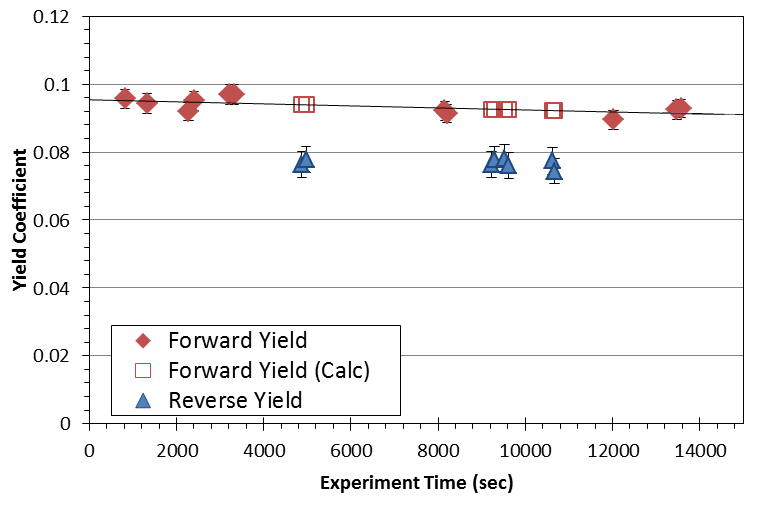}
  \caption[Comparison of Forward and Reverse Yield]{Comparison of the forward and reverse photoproduct Yield, excited by an excitation pulse centered at $\lambda_0=620$ nm with $P_{Ex}=250\mu$W. The average forward yield is $Y_{R,FR}$=0.093$\pm$0.002 while the average reverse reaction yield is $Y_{FR,R}$=0.077$\pm$0.0015 (82\% of $Y_{R,FR}$). This corresponds to a photoequilibrium population of (\pr, \pfr)=(0.45,0.55).}
  \label{YR_YF}
\end{figure}

The excitation spectrum, generated by spectrally filtering the broadband NOPA beam between 600 nm to 640 nm, was chosen to excite both the $P_{R}$ and $P_{FR}$ forms of the Cph8 protein, initiating forward and reverse photoswitching reactions.
Measurements comparing the forward and reverse Yield coefficient are presented in Figure~\ref{YR_YF}, hollow squares mark the values of the forward Yield used to calculate the reverse Yield. The excitation power is 250 $\mu$W and the flow rate setting is q=125 on the pump driver. The average forward Yield is $Y_{R,FR}$=0.093$\pm$0.002 while the average reverse Yield is $Y_{FR,R}$=0.077$\pm$0.0015 (82\% of $Y_{R,FR}$). Measurements across multiple days show a consistent ratio between the forward and reverse product Yield coefficient, the reverse Yield $Y_{FR,R}$ is typically 80-85\% of the forward Yield $Y_{R,FR}$.
This ratio of forward and reverse Yield coefficients produces a photoequilibrium \pfr population of $[P_{FR}^{(eq)}= 0.55\pm .033$. This equilibrium population is lower than the maximum photoequilibrium threshold of $P_{FR}\simeq 0.65$ achieved by linear excitation by narrow-bandwidth filtered light at 655 nm because the excitation pulse is spectrally broader and centered at a higher frequency than the optimal linear exposure.
This result is intentional, to produce a wavepacket of vibrational states with extra energy on the excited state potential energy surface, rather than exciting at the (lower energy) peak of the \pr absorption spectrum.

\begin{figure}
  \centering
  \includegraphics[width=6in]{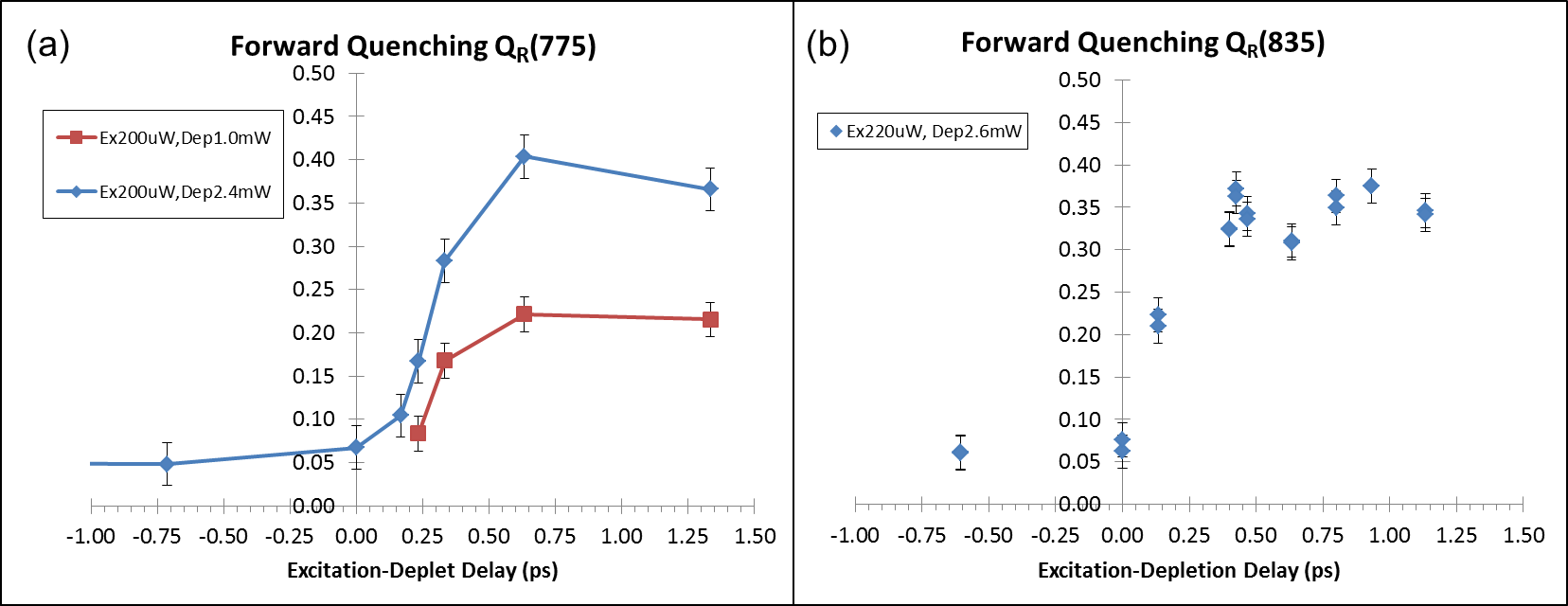}
  \caption[$Q_R$ Stimulated Depletion Quenching of Forward Reaction]{Forward stimulated depletion quenching coefficients using a depletion pulse with 10 nm spectral width centered at (a) 775 nm and (b) 835 nm.
   Measurements of $Q_{R}$ were taken over a series of Excitation-Depletion delay times and at multiple powers (labeled in Key).
   The Quenching coefficient has a fast rise to peak depletion characteristic of the ultrafast dynamics of the Cph8 molecule.
  }
  \label{FigQR}
\end{figure}

Knowing the un-depleted, linear Yield coefficients for the forward and reverse reactions, we move to characterizing the Quenching coefficient.  The dependence of the forward Quenching coefficient on the delay between the excitation and depletion pulses is shown Figure~\ref{FigQR} for a depletion wavelength of $\lambda_{Dep}$=775 nm (a) and $\lambda_{Dep}=835$ nm (b). We observe the sub-picosecond rise time to peak depletion, corresponding to the expected ultrafast excited state dynamics of the Cph8 switch.
The depletion curves for two depletion pulse powers are shown in Figure~\ref{FigQR}(a), and we see that the Quenching coefficient doubles when the average power of the depletion pulse is doubled, as expected.

%

\begin{figure}
  \centering
  \includegraphics[width=\linewidth]{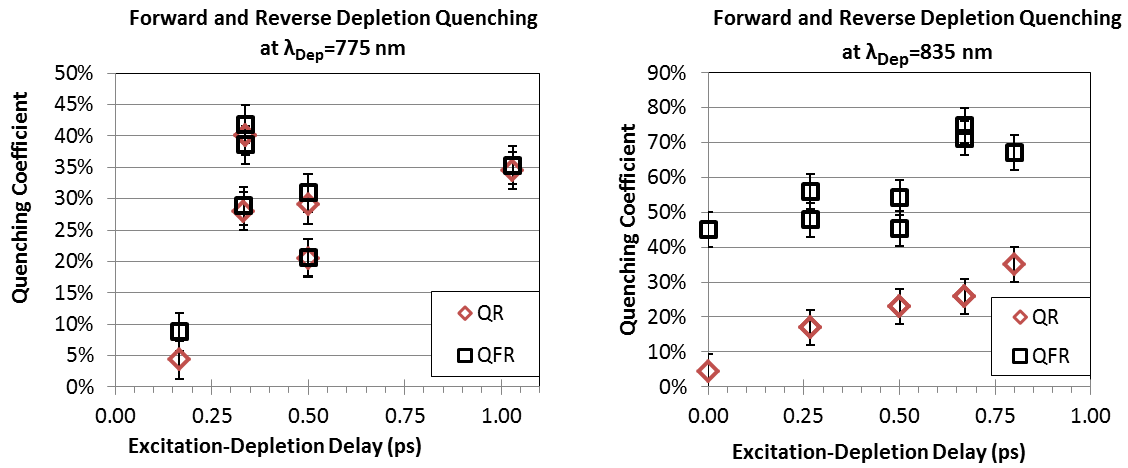}
  \caption[Comparison of Quenching at Long and Short Depletion Wavelengths]{Comparing Quenching coefficients of the forward ($Q_{R}$) and reverse ($Q_{FR}$) photoswitching reaction at long and short depletion wavelengths. Measurements are plotted with respect to pulse delay times, all measurements taken at excitation power $P_{Ex}=200\, \mu W$ and depletion power $P_{Dep}=2.0\, mW$. For $\lambda_{Dep}=775$ nm there is little selectivity (ratio $\nicefrac{Q_{FR}}{Q_{R}}\simeq$1), while the $\lambda_{Dep}=835$ nm depletion pulse produces consistently higher quenching coefficient in the reverse direction compared to the forward, with the ratio increasing at short delay times. Both these results are in qualitative agreement with recent simulations of \cite{QuineTheory2018}.}\label{QRvsQF}
\end{figure}

The calculation of the reverse Quenching coefficient requires knowledge of all five other \emph{characteristic coefficients} (see Table~\ref{Tab:CharCoeff}), making it the most difficult to accurately measure. Because of the numerous measurements needed to calculate this final reaction coefficient, we were only able to measure a Full Data Series at a few control parameter settings before the sample degradation made the measurements quantitatively inconsistent.
However, there are clear repeatable trends in the data that can be consistently measured over many experiments which is consistent with the simulations of \cite{QuineTheory2018}.
In Figure~\ref{QRvsQF} we compare the depletion Quenching coefficient in the forward and reverse photoswitching reactions depleted by pulses centered at $\lambda_{Dep}=775$ nm (a) and $835$ nm (b).
Measurements for both graphs were taken with an excitation pulse power of 200 $\mu$W, a depletion pulse power of 2.0 mW, an excitation spectrum of 30 nm centered at 625 nm and a depletion pulse spectral width of 10 nm.
When the depletion pulse spectrum is centered at 775 nm, the forward and reverse Quenching coefficients are nearly identical for all delays, giving an average ratio of ${Q_{FR}}/{Q_{R}}=1.02\pm0.03$.
As predicted by the simulations in \cite{QuineTheory2018}, depletion at this wavelength grants no selective control over the reaction.
When the depletion pulse spectrum is centered at 835 nm, there is significant enhancement of the Quenching coefficient of the reverse reaction, with an average ${Q_{FR}}/{Q_{R}}$ ratio of 2.1$\pm$0.4 for excitation-depletion delays between 0.2-0.8 ps, with a much higher ratio of 10.0 at $t_{Ex,Dp}=0.00$ ps.
The value of the reverse Quenching coefficient being at least double the forward Quenching coefficient and ${Q_{FR}}$ remaining high while ${Q_{R}}$ falls off at short excitation-depletion delays is in good agreement with the predictions of modeling \cite{QuineTheory2018}, however the quenching effect persists to much longer times than predicted by the simulations.
The spectral dependence and power dependence of the measured data are in better agreement with the model than the temporal dependence. This is reasonable, considering that the simulation used a kinetic model which is a simple approximation of the molecule's true ultrafast coherent dynamics.

\begin{figure}
  \centering
  \includegraphics[width=4.5in]{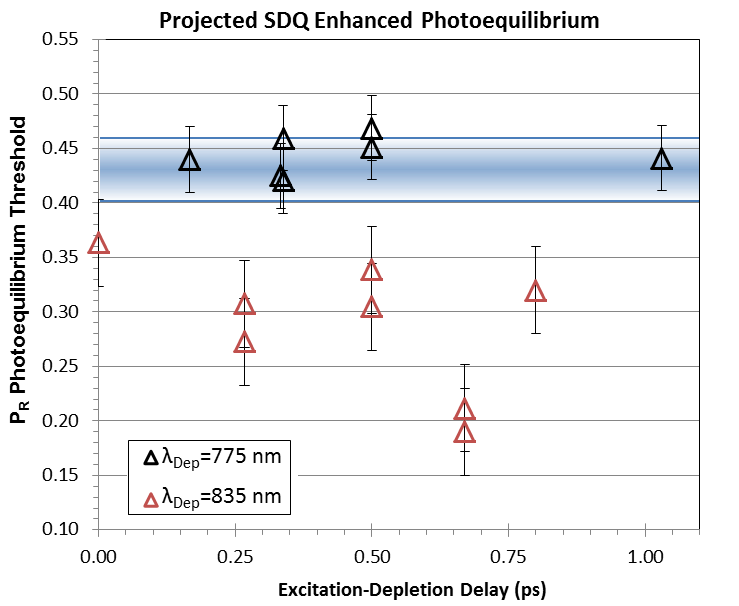}
  \caption[Projected \pfr Equilibrium Population]{Final \pr product state populations at the SDQ enhanced photoequilibrium. Data from two depletion wavelengths are shown plotted with respect to the Excitation-Depletion pulse delay. Longer depletion wavelength and sub-picosecond delays consistently produce significant reduction of the final \pr population over the linear photoswitching equilibrium threshold (shown by the blue highlighted region at 0.45), while shorter depletion wavelength produces no change in the final photoequilibrium population. The maximum measured result is $[P_R, P_{FR}]_{EQ}^{SDQ}=[19\%, 81\%]$ at $\lambda_{Dep}=835$ nm, $P_{Ex}=200$ $\mu$W, $P_{Dep}=2.2$ mW, and $t_{Ex,Dep}=0.66$ ps.}\label{Fig:OG:ProjPFReq}
\end{figure}

The enhanced depletion quenching of the reverse, $P_{FR}\rightarrow P_{R}$, photoswitching reaction produces a shift in the photoequilibrium: building up a greater final product population of $P_{FR}$ and removing the undesired $P_{R}$ population over many laser pulse iterations.
To estimate this enhanced equilibrium we calculate the eigensolutions of the OTM constructed from the measured forward and reverse Yield and Quenching coefficients Eq~\ref{eq:OG:expOTMccs}.
The errors on the photoequilibrium are obtained by statistical analysis of the calculated eigensolutions of 1000 matrixes constructed from the measured characteristic coefficients varied by a Gaussian noise term scaled by the magnitude of the associated coefficient's error.
The SDQ enhanced switching associated with the measurements in Figure~\ref{QRvsQF} are plotted in Figure~\ref{Fig:OG:ProjPFReq} for the depletion pulses centered at 835 nm and 775 nm. The improved performance manifest itself in the reduction of the undesirable $P_{R}$ state.

The linear exposure photoequilibrium is calculated by omitting the Quenching coefficients from the OTM, which gives $[P_{R}]_{EQ}^{Lin}=0.45\pm$0.033. This is indicated in Figure~\ref{Fig:OG:ProjPFReq} by the blue highlighted region.
The depletion at $\lambda_{Dep}=775$ produces no shift away from this limit, however all exposures to $\lambda_{Dep}=835$ produce significant reduction below the linear threshold.
The highest ratio of depletion quenching coefficients at $t_{Ex,Dp}=0$ ps does not actually produce the highest fidelity equilibrium population as, the greater reverse quenching coefficients at longer times has more impact than the forward quenching coefficient going to zero at short delays.
The strongest reduction of spectral cross talk $[P_{R}]_{EQ}^{SDQ}=0.19$ is achieved for the parameter set
$\left[\langle P_{Ex}\rangle, \lambda_0^{Ex}, \Delta\lambda_{Ex}, \langle P_{Dp}\rangle, \lambda_0^{Dp},\Delta\lambda_{Dp}, t_{Ex,Dp}\right]=[200 \mu W, 625 nm, 30 nm, 2.2 mW, 835 nm, 10 nm, 0.667 ps]$, maximizing the functional dynamic range of the Cph8 switch using SDQ of the photoswitching reaction.



\section{Conclusions}\label{Sec:OG:Conclusions}


The results presented in this paper demonstrate the capability to exploit stimulated depletion of the $P_{FR}$ chromophore excited state to quench the photoisomerization reaction before it takes place, diminishing the unwanted reverse photoswitching of the Cph8 optogenetic switch.
Further, by choosing particular spectral and temporal characteristics of the excitation and depletion pulses, it is possible to selectively deplete the different chromophore states by exploiting their unique excited state dynamics and spectral response.
The SDQ mechanism is proven feasible without pulse shaping, though the added capability of pulse shaping should enable even greater control of the Cph8 switch (especially at shorter excitation-depletion delays) and open up the prospect of discrimination and orthogonal control of multiple optogenetic switches in future experiments.
Selective stimulated depletion of a chosen chromophore excited state thereby granting control over the quenching of different photo-reaction pathways, should be able to alter the final product state distributions of the photoswitching reaction. For Cph8 depletion at longer wavelengths (peaking at 835 nm) and short excitation-depletion delay times (peaking at $\sim$0.6 ps) results in reverse photo-reaction quenching coefficient that is more than twice as strong as the forward quenching coefficient giving a final enhanced equilibrium threshold population of $[P_{R},P_{FR}]_{EQ}^{SDQ}=[0.19, 0.81]$. This is a 58\% reduction of the unwanted \pr state compared to the excitation pulse alone and 46\% reduction compared to the best linear photoequilibrium of $[P_{R},P_{FR}]_{EQ}^{Linear}=[0.35, 0.65]$.

We have shown that photo-damage is not a significant impediment to nonlinear control of photoswitching. There is a large window of laser intensities where it is possible to drive nonlinear interactions without observing significant photo-damage of fresh samples. We have also shown that the sub-picosecond dynamics of the $P_{FR}$ isomerization was instrumental in enabling selective quenching of the forward and reverse photo-reaction.

A recent work developed a simple flexible model of the Cph8 photoswitching system that can be adapted to the simulation of a number of alternate optogenetic switching systems or simultaneous control of multiple optogenetic switches \cite{QuineTheory2018}.
The model is based on incoherent rate equation kinetics of the two switch states, and incorporates experimentally measured optical and dynamical characteristics of the molecular switch to simulate it's optically driven dynamics.
The model reproduces the expected steady-state photoequilibrium population distribution under saturated linear illumination with red or far-red light.
The predictions of this model are found to be in qualitative agreement with the measurements in the present paper, and the simulations predict a maximum enhancement of the steady-state equilibrium concentration of 93\% $P_{FR}$ and 7\% $P_{R}$. The largest discrepancies associated with differences in the excitation-depletion time dependence of the SDQ photoequilibrium enhancement. This difference is likely due to the omission of coherence effects from the simulations, as the coherent excited state wave packet dynamics of the molecule are sure to play a role in the effectiveness of the SDQ mechanism, especially at short delay times and eventually of even more importance with pulse shaping.
The model and experiments agree well in the basic aspects of the SDQ mechanism with regard spectral characteristics.
The un-depleted reaction yields in the forward and reverse directions are linearly related to the power of the excitation pulse, and the reverse yield is approximately 70\% of the forward yield when excited by a pulse spectrum of 40 nm width centered at 625 nm, in agreement with the simulations \cite{QuineTheory2018} and the literature \cite{Lamparter2001}.
Depletion at shorter wavelengths ($\lambda_{Dep}$ = 775 nm) quenches both photoswitching reaction directions with comparable strength and has little effect on the equilibrium concentrations but greatly slows the reaction. This was confirmed at multiple excitation and depletion laser powers and delays.
Depletion at 835 nm and sub-picosecond excitation-depletion delays increases the relative enhancement of the reverse quenching in comparison to the forward, with the ratio increasing as the excitation-depletion pulse delay was reduced.
Our measurements showed that for $\lambda_{Dep}=$835 nm and an excitation-depletion delay between 0.2-0.8 ps the ratio of reverse to the forward quenching coefficients had an average of $\langle\nicefrac{Q_{FR}}{Q_{R}}\rangle=2.1 \pm 0.4$, increasing as the delay is decreased to  $\langle\nicefrac{Q_{FR}}{Q_{R}}\rangle=4.7$ at $\uptau_{Ex,Dep}$= 0.066 ps and  $\langle\nicefrac{Q_{FR}}{Q_{R}}\rangle=10.0$ at $\uptau_{Ex,Dep}$= 0.0 ps.

With the current experimental setup we were unable to expose the sample to sufficient laser iterations to enable the SDQ-enhanced photoequilibrium to overcome the slower reaction rate.
However, with the iterative mapping in Eq. 4, we were able to calculate the SDQ-enhanced photoequilibrium threshold by computing the eigensolutions to the OTM composed of the experimentally measured Characteristic Coefficients. Alternative experimental design should enable a full set of laboratory iterations

The demonstrated capability to selectively quench the photo-isomerization of the chromophore in the Cph8 switch is a significant step toward complete coherent control of optogenetic switches, enabling full dynamic range activation/deactivation of cellular signaling pathways and discriminating control of multiple switches with overlapping action spectra allowing multiplexed control of several switches simultaneously.


\singlespacing

\bibliographystyle{unsrt}
\bibliography{bib_thesis}
\end{document}